\documentclass[sigconf]{acmart}

\AtBeginDocument{%
  }

\setcopyright{cc}
\setcctype[4.0]{by}
\acmDOI{}
\acmISBN{}
\acmConference[LOCO 2024]{1st International Workshop on Low Carbon Computing}{December 3, 2024}{Glasgow/online}

\usepackage{url}
\usepackage{dsfont}
\usepackage{algorithm}
\usepackage{algpseudocode}
\newcommand{\carbonopt}[0]{Carbon-Only Optimization }
\newcommand{\wateropt}[0]{Water-Only Optimization }
\newcommand{\landuseopt}[0]{Land-Use-Only Optimization }
\newcommand{\preferencebased}[0]{Preference-based Optimization }
\usepackage{subcaption}
\usepackage{graphicx}
\usepackage{caption}
\begin{document}

\title{Environmentally-Conscious Cloud Orchestration Considering Geo-Distributed Data Centers
%A Multidimensional Metric for Holistic Sustainability in Computing Systems/Services
%Accounting for multidimensional impact of computing services 
}

\author{Giulio Attenni, Novella Bartolini}
\affiliation{%
\{attenni,bartolini\}@di.uniroma1.it\\
  \institution{"La Sapienza" University of Rome}
  \country{Italy}
}



\begin{abstract}
This paper presents a {theoretical discussion} for environmentally-conscious job deployment and migration in cloud environments, aiming to minimize the environmental impact of resource provisioning while incorporating sustainability requirements. As the demand for sustainable cloud services grows, it is crucial for cloud customers to select data center operators based on sustainability metrics and to accurately report the ecological footprint of their services. To this end, we analyze sustainability reports and define comprehensive environmental impact profiles for data centers, incorporating key sustainability indicators. We formalize the problem as a{n} optimization model, balancing multiple environmental factors while respecting user preferences. A simulative case study demonstrates the {potential} of our approach compared to baseline strategies that optimize for single sustainability factors. 
\end{abstract}




\maketitle

\section{Introduction}
The demand for cloud services is expected to increase in the coming years, and the associated environmental impact represents a critical concern. Data center sustainability is gaining more and more attention and the need for comprehensive measures to reduce the environmental impact is becoming clear. In response to the growth of the data center economy, the European regulatory landscape rapidly evolved. 
The JRC \cite{jrc} in 2018 established the voluntary European Code of Conduct for Data Centers \cite{EU_Code_of_Conduct}.
Then, in 2021, the Climate Neutral Data Center Pact was launched, to make data centers and cloud infrastructure services in Europe climate-neutral by 2030. Signatories pledge to meet quantifiable goals, such as using 100\% renewable energy while also making recycling and water conservation a priority \cite{Climate_Neutral_Data_Centre_Pact}.
Moreover, from 2024 the Energy Efficiency Directive sets reporting obligations for data centers with a power demand of at least 500kW \cite{EU_Energy_Efficiency_Directive}.
To enhance and standardize sustainability reporting among companies in 2023, the Corporate Sustainability Reporting Directive %(CSRD)
has been enacted, which imposes that companies must disclose detailed information on their environmental and social impacts, sustainability risks, and governance practices \cite{EU_Corporate_Sustainability_Reporting}.
Finally, in March 2024 the EU Commission  adopted a new delegated regulation to establish an EU-wide scheme to rate the sustainability of data centers \cite{EU_Sustainability_Rating_Scheme}.
Thus, both for data center operators and companies relying on cloud-intensive computations that need to meet their environmental pledges, it is crucial to consider the environmental footprint of their operations in corporate sustainability reports.
\paragraph{Contributions}
This paper presents the following contributions:
\begin{itemize}    
    \item Analysis of sustainability reports, extracting meaningful environmental indicators to inform decision-making.
    \item Definition of sustainability profiles and formalized metrics to assess the sustainability of data centers.
    \item Proposal of an environmentally-conscious cloud orchestrator that optimizes job provisioning and migration to minimize the environmental footprint, taking into account users' requirements and the need to comply with sustainability standards for data centers.
    \item Simulative case study to demonstrate the potential of our sustainability-aware orchestration by comparing our approach with some baseline approaches.
\end{itemize}
\paragraph{Paper Organization}
The rest of this paper is structured as follows: Section \ref{sec:background} provides background on data center sustainability, sustainability report analysis, sustainability profiles, and formal metric definitions. Section \ref{sec:problem} introduces the cloud orchestration problem. Section \ref{sec:experiments} details our experiments and presents the results. Section \ref{sec:relwork} discusses related work in sustainable cloud orchestration. Finally, Section \ref{sec:conclusions} concludes the paper {highlighting the threats to the validity of our work, }and outlines future research directions.

\section{Background on Data Centers' Environmental Impact}
In this section, we discuss the environmental impact of data centers, focusing on key factors such as energy consumption, carbon emissions, water usage, and land occupation. We provide an analysis of data extracted from sustainability reports of major cloud providers, i.e. Google, Meta, and Azure. Finally, a formalization of sustainability profiles is presented.
\label{sec:background}
\subsection{Impact factors} 
Data centers' environmental impact can be categorized into two broad areas: operational and construction/dismantling impacts. Each phase of a data center's life cycle contributes to its ecological footprint.

\paragraph{Operational Impact}
 Several metrics can be considered to account for operational impact \cite{SunbirdDCIM_Sustainability_Metrics}.
The impact of energy production has to be considered, since the ongoing operation of data centers is resource-intensive, particularly in terms of energy consumption and water usage. Data centers are among the most energy-demanding facilities globally. They require large amounts of electricity to power servers, cooling systems, and backup power supplies. It is estimated that data centers in 2022 accounted for nearly 500 TWh, and it is expected to rise above 800 TWh of consumption in 2026, with cloud-based services, 5G networks, AI, and cryptocurrencies being the major drivers of such demand increase \cite{IEA2024report}. This energy usage results in significant GreenHouse Gas (GHG) emissions, especially in regions where electricity is generated from fossil fuels.
In fact, the environmental impact of data centers is directly tied to the energy mix of their respective grids.
However, GHG emissions are not the only aspect that characterizes the environmental footprint of data centers. Namely: \textit{water footprint}, which contributes to environmental pollution; \textit{land use}\cite{OurWorldInData_Land_Use_Energy_Sources}, which affects ecosystems and biodiversity; \textit{deathprint} \cite{Conca_Energys_Deathprint_2012,OurWorldInData_Safest_Energy_Sources}, which is the number of people killed per kWh produced; \textit{waste}, which contributes to environmental pollution; \textit{critical raw metals} and \textit{material use}, which affect resource depletion. 
Cooling systems have a huge impact on the environment since they are both energy-intensive and water-intensive. Indeed, data centers rely on vast amounts of water for cooling purposes. It is estimated that a typical mid-sized data center can use approximately 25.5 million liters of water each year when employing traditional cooling methods \cite{H2O_Data_Center_Water_Use}. This heavy water reliance can exacerbate local water scarcity, especially in regions prone to drought. Examples of metrics related to cooling systems are: the Air Economizer Utilization Factor %, the Water Economizer Utilization Factor, 
and the Water Usage Effectiveness \cite{SunbirdDCIM_Sustainability_Metrics}.

\paragraph{Construction and Dismantling Impact}
Data center facilities also have a significant environmental impact for what concerns the construction and eventual dismantling.
Large amounts of raw materials like concrete, steel, copper, and rare earth elements are required for their construction. Moreover, these facilities also occupy significant land.
The level of the ecological footprint of a data center building can be formally certified, e.g. with the LEED certification scheme  \cite{NetZero_Sustainable_Data_Center_Standards}.
{Moreover, in the context of data centers, it is important to consider not only building materials but also the IT equipment such as servers, networking devices, and storage systems. The production and end-of-life disposal of such hardware contributes significantly to environmental impact. \cite{embodiedhw}
Furthermore,} the dismantling of data centers generates substantial e-waste, which often contains toxic materials{, like lead, nickel, and mercury, which,} if not properly recycled, these can leach into soil and water, ultimately having an adverse impact on human health and the environment \cite{pinto2008waste}.

\subsection{Sustainability Reports Analysis}
Sustainability reports published by data center operators provide valuable insights into their environmental impact. These reports typically include key metrics related to energy consumption, carbon emissions, water usage, and waste management, including key metrics such as Power Usage Effectiveness (PUE), Water Usage Effectiveness (WUE), or water withdrawal. 
{These metrics are respectively defined as the ratio between the facility's total energy consumption and energy consumption due to the IT equipment, and between the total water withdrawal and energy consumption due to the IT equipment.}

Azure \cite{Azure_Iowa}, 
\cite{Azure_Italy, Azure_Texas, Azure_Sweden, Azure_Arizona, Azure_Illinois, Azure_Ireland, Azure_Wyoming, Azure_Singapore, Azure_Washington, Azure_Netherlands} provides both PUE and WUE metrics, which offer insight into the energy and water efficiency of its data centers. Despite this level of reporting, to the best of our knowledge, there is no publicly available data on the actual power consumption or total water withdrawal at data center or regional level. This lack of absolute consumption figures makes it difficult to assess the total environmental footprint of Azure's facilities and compare them directly with other providers that report different sustainability metrics, such as Google’s facility-level water withdrawal \cite{Google2024EnvironmentalReport} or Meta’s facility-level energy consumption \cite{Meta2024SustainabilityReport}. 

In Tables \ref{tab:Azure_report}, \ref{tab:google_report} and \ref{tab:meta_report} we summarize information found in official sustainability reports and fact sheets for what concern energy and water usage. Furthermore, land-related data reflect the total property size of the data center including auxiliary infrastructures, and are sourced from Datacenter.com \cite{Datacenters_website}, a global database that aggregates information on data center properties.

The information about Azure's data centers is shown in Table \ref{tab:Azure_report}. Azure's data centers exhibit varying levels of efficiency, with PUE values ranging from 1.12 (Italy) to 1.358 (Singapore). WUE fluctuates significantly, from as low as 0.023 l/kWh in Italy to 2.24 l/kWh in Arizona, highlighting regional dependencies. Land occupation varies widely, and related information is not available for all data centers. 

Google’s sustainability data reported in Table \ref{tab:google_report} presents a different perspective, as the company reports water withdrawal instead of WUE. Water usage varies considerably across locations, with Oklahoma withdrawing the highest volume (3.9 billion liters) and Ireland the lowest (2 million liters). PUE values are relatively consistent across its facilities, ranging from 1.07 (Oregon) to 1.13 (Texas). However, direct comparison with Azure’s data is difficult, as Google does not disclose IT power consumption at the data center level, making it impossible to directly calculate WUE.

{In Table \ref{tab:aws_report}, we report PUE values provided by Amazon concerning AWS regions, which span from 1.08 (Melburne) and 1.5 (Hyderabad).  
Amazon does not provide any information on water withdrawal or energy consumption. However, it does provide a global annual WUE value (0.18).}

Meta’s data, reported in Table \ref{tab:meta_report}, introduce another challenge in comparability. Unlike Azure and Google, Meta provides only a global PUE measure (1.08) rather than specific values per data center. 
Additionally, Meta includes facility energy consumption figures, but without a breakdown of IT load, direct comparison with other providers remains challenging.

{Gathering data, we noticed that most of the companies declare to have a high percentage of recycled waste or even to have reached the goal of zero waste. However, we could not find useful data for comparison at the level of data centers.}

In general, due to differences in reporting methodologies, direct cross-company comparisons require additional calculations and approximations. The lack of uniformity in data disclosure means that sustainability footprints cannot be directly compared.

{\footnotesize
\begin{table}[]
    \centering
    \begin{tabular}{|c|c|c|c|c|}
    \hline
        \textbf{Location} & \textbf{PUE} & \textbf{WUE ($l/kWh$)} & \textbf{Land ($m^2$)} & \textbf{Source}\\ \hline
        Arizona & 1.223 & 2.24 & 22730& \cite{Azure_Arizona, Azure_Arizona_space}\\\hline
        Illinois&1.346&0.79&65032&\cite{Azure_Illinois, Azure_Illinois_space}\\\hline
        Iowa&1.16&0.19&37904&\cite{Azure_Iowa, Azure_Iowa_space}\\\hline
        Texas&1.307&1.82&43664&\cite{Azure_Texas, Azure_Texas_space}\\\hline
        Washington&1.156&1.09(2021)&74322&\cite{Azure_Washington, Azure_Washington_space}\\\hline
        Wyoming&1.125&0.23& - &\cite{Azure_Wyoming}\\\hline

        Ireland & 1.197(2021) & 0.03(2021) & 28149 & \cite{Azure_Ireland, Azure_Ireland_space}\\\hline
        Italy&1.12&0.023&-&\cite{Azure_Italy}\\\hline
        Netherlands&1.158&0.08&-&\cite{Azure_Netherlands}\\\hline
        Sweden&1.172&0.16&-&\cite{Azure_Sweden}\\\hline
        Singapore&1.358&2.06&32516&\cite{Azure_Singapore, Azure_Singapore_space}\\\hline
    \end{tabular}
    \caption{Azure Data Centers reported data (PUE and WUE referred to 2022 or 2021 when specified)}
    \label{tab:Azure_report}
\end{table}
}
{\footnotesize
\begin{table}[]
    \centering
    \begin{tabular}{|c|c|c|c|c|}
\hline
\textbf{Location} & \textbf{PUE} & \textbf{H2OWdr ($Ml$)} & \textbf{Land ($m^2$)} & \textbf{Source}\\\hline
Alabama & 1.1 & 604.91 & 25084 & \cite{Google2024EnvironmentalReport, Google_AL}\\\hline
Georgia & 1.09 & 1585.33 & 120774 & \cite{Google2024EnvironmentalReport, Google_GA}\\\hline
Nebraska & 1.09 & 621.56 & - & \cite{Google2024EnvironmentalReport}\\\hline
Nevada (Henderson) & 1.08 & 1036.45 & 6968 & \cite{Google2024EnvironmentalReport, Google_NV}\\\hline
Nevada (Storey County) & 1.19 & 7.19 & - & \cite{Google2024EnvironmentalReport}\\\hline
North Carolina & 1.09 & 1355.56 & 31308 & \cite{Google2024EnvironmentalReport, Google_NC}\\\hline
Ohio & 1.1 & 575.38 & 25548 & \cite{Google2024EnvironmentalReport, Google_OH}\\\hline
Oklahoma & 1.1 & 3925.85 & 130064 & \cite{Google2024EnvironmentalReport}\\\hline
Oregon & 1.07 & 726.23 & - & \cite{Google2024EnvironmentalReport}\\\hline
Texas & 1.13 & 621.94 & 74322 & \cite{Google2024EnvironmentalReport, Google_TX}\\\hline
South Carolina & 1.1 & 3206.99 & - & \cite{Google2024EnvironmentalReport}\\\hline
Tennessee & 1.1 & 1294.61 & - & \cite{Google2024EnvironmentalReport}\\\hline
Chile & 1.09 & 721.88 & - & \cite{Google2024EnvironmentalReport}\\\hline
Belgium & 1.09 & 1320.73 & 83613 & \cite{Google2024EnvironmentalReport, Google_BE}\\\hline
Denmark & 1.1 & 102.21 & 13471 & \cite{Google2024EnvironmentalReport, Google_DK}\\\hline
Finland & 1.09 & 11.36 & 16000 & \cite{Google2024EnvironmentalReport, Google_FI}\\\hline
Ireland & 1.08 & 2.27 & 28800 & \cite{Google2024EnvironmentalReport, Google_IE}\\\hline
Netherlands & 1.08 & 1121.99 & - & \cite{Google2024EnvironmentalReport}\\\hline

    \end{tabular}
    \caption{Google Data Centers reported data (PUE and water withdrawal referred to 2023, H2OWdr converted to megaliters)}
    \label{tab:google_report}
\end{table}
}

{\footnotesize
\begin{table}[]
    \centering
    \begin{tabular}{|c|c|c|c|}
    \hline
\textbf{Location}&\textbf{PUE}&\textbf{Land ($m^2$)}&\textbf{Source}\\\hline
Canada & 1.22 & - & \cite{AmazonAWS_Efficiency}\\ \hline
California & 1.17 & 3326 & \cite{AmazonAWS_Efficiency, california_aws}\\ \hline
Northern Virginia & 1.15 & - & \cite{AmazonAWS_Efficiency}\\ \hline
Ohio & 1.12 & 102007 & \cite{AmazonAWS_Efficiency,datacenters2024dublin,datacenters2024hayden5113,datacenters2024newalbany,datacenters2024hayden5117,datacenters2024jugbeach}\\ \hline
Oregon & 1.13 & - & \cite{AmazonAWS_Efficiency}\\ \hline
Brasil & 1.18 & - & \cite{AmazonAWS_Efficiency}\\ \hline
Ireland & 1.1 & 2159 & \cite{AmazonAWS_Efficiency,datacenters2024dub9,datacenters2024dub10}\\ \hline
Sweden & 1.12 & 13471 & \cite{AmazonAWS_Efficiency,datacenters2024katrineholm}\\ \hline
Germany & 1.33 & - & \cite{AmazonAWS_Efficiency}\\ \hline
Spain & 1.11 & - & \cite{AmazonAWS_Efficiency}\\ \hline
South Africa & 1.24 & - & \cite{AmazonAWS_Efficiency}\\ \hline
Bahrain & 1.32 & - & \cite{AmazonAWS_Efficiency}\\ \hline
UAE & 1.36 & - & \cite{AmazonAWS_Efficiency}\\ \hline
India (Hyderabad)& 1.5 & - & \cite{AmazonAWS_Efficiency}\\ \hline
India (Mumbai)& 1.44 & - & \cite{AmazonAWS_Efficiency}\\ \hline
China & 1.26 & - & \cite{AmazonAWS_Efficiency}\\ \hline
Singapore & 1.3 & - & \cite{AmazonAWS_Efficiency}\\ \hline
Indonesia & 1.35 & - & \cite{AmazonAWS_Efficiency}\\ \hline
Japan & 1.32 & - & \cite{AmazonAWS_Efficiency}\\ \hline
Australia (Melbourne) & 1.08 & - & \cite{AmazonAWS_Efficiency}\\ \hline
Australia (Sydney) & 1.15 & 21367.69 & \cite{AmazonAWS_Efficiency,datacenters2024syd51, datacenters2024syd52}\\ \hline
    \end{tabular}
    \caption{AWS Reported data (WUE = 0.18 \cite{amazon2024water}, PUE referred to 2023)}
    \label{tab:aws_report}
\end{table}
}

{\footnotesize
\begin{table}[]
    \centering
    \begin{tabular}{|c|c|c|c|c|}
    \hline
\textbf{Location}&\textbf{F. cons. ($MWh$)}&\textbf{H2OWdr ($Ml$)}&\textbf{Land ($m^2$)}&\textbf{Source}\\\hline
Alabama&614,198&152&90116&\cite{Meta2024SustainabilityReport,Meta_Alabama}\\\hline
Georgia&968,565&61&-&\cite{Meta2024SustainabilityReport}\\\hline
Illinois&138,965&55&-&\cite{Meta2024SustainabilityReport}\\\hline
Iowa&1,243,306&173&133780&\cite{Meta2024SustainabilityReport,Meta_Iowa}\\\hline
Nebraska&1,148,091&123&2415478&\cite{Meta2024SustainabilityReport,Meta_Nebraska}\\\hline
New Mexico&1,110,100&283&37161&\cite{Meta2024SustainabilityReport,Meta_NewMexico}\\\hline
North Carolina&507,068&55&68748&\cite{Meta2024SustainabilityReport,Meta_NorthCarolina}\\\hline
Ohio&793,063&72&55742&\cite{Meta2024SustainabilityReport,Meta_Ohio}\\\hline
Oregon&1,375,321&180&297290&\cite{Meta2024SustainabilityReport,Meta_Oregon}\\\hline
Tennessee&116,520&3&-&\cite{Meta2024SustainabilityReport}\\\hline
Texas&1,029,570&404&2043866&\cite{Meta2024SustainabilityReport, Meta_Texas}\\\hline
Utah&787,740&87&90116&\cite{Meta2024SustainabilityReport,Meta_Utah}\\\hline
Virginia&805,061&42&41806&\cite{Meta2024SustainabilityReport,Meta_Virginia}\\\hline
Denmark&518,005&371&25084&\cite{Meta2024SustainabilityReport,Meta_Denmark}\\\hline
Ireland&953,837&659&86000&\cite{Meta2024SustainabilityReport,Meta_Ireland}\\\hline
Sweden&351,931&50&92903&\cite{Meta2024SustainabilityReport,Meta_Sweden}\\\hline
    \end{tabular}
    \caption{Meta Data Centers reported data (PUE = 1.08, Facility consumption and water withdrawal referred to 2023)}
    \label{tab:meta_report}
\end{table}
}

\subsection{Sustainability Profile}
\label{sec:profiles}
Based on the reported data, we define sustainability profiles to evaluate data centers according to their environmental performance.
These profiles take into account mainly four main factors: carbon emissions due to their power consumption; water footprint due to both their power consumption and cooling needs; and land use due to both their power consumption and the space occupied by the data center property; and the non-recycled/non-repurposed electronic waste.
Table \ref{tab:sustainability_profile} summarizes the sustainability profile of a data center.

Let us now formally define the metrics for the sustainability profiles. Let us consider a data center $d$ located in a specific region $R_d$.

\paragraph{Carbon footprint} Each data center $d$ is characterized by its PUE denoted as $PUE_d$. We note that data centers may have on-site power plants to meet part of their demands \cite{8511065,10.1145/1993744.1993791}. The carbon intensity of a data center, $CI_d$ \cite{CI}, depends on its energy source mix, with a fraction $P_d$ of its power coming from an on-site power grid and the remaining $1 - P_d = P_{R_d}$ sourced from the regional grid, which has a carbon intensity $CI_{R_d}$. The total carbon emissions generated by executing a job $j$ consuming $E_j$ kWh in data center $d$ is computed as:

\begin{equation}
    (E_j \cdot PUE_d) \cdot (CI_d \cdot P_d + CI_{R_d} \cdot P_{R_d} )
\end{equation}

\paragraph{Water footprint} Each data center has a metric $WUE_d$, measuring water consumed per unit of energy. As proposed by Jiang et al. \cite{jiang2025waterwise} the Water Scarcity Factor (WSF) of the regions can be used to address water stress scenarios,  thus let $WSF_{R_d}$ be the WSF of the region $R_d$. The on-site water footprint for executing a job is:

\begin{equation}
    E_j \cdot WUE_d \cdot (1 + WSF_{R_d})
\end{equation}

Additionally, incorporating on-site power grid contributions, the on-site water footprint is refined as:

\begin{equation}
    E_j \cdot (WUE_d + P_d \cdot EWIF_d) \cdot (1 + WSF_{R_d})
\end{equation}

The off-site water footprint, arising from energy sourced from the regional grid is also defined similarly.
$EWIF_{R_d}$ is the Energy Water Intensity Factor of $R_d$ which accounts for the water footprint due to the power generation.
The on-site water footprint for executing a job is expressed as:

\begin{equation}
    (E_j \cdot PUE_d) \cdot (P_{R_d} \cdot EWIF_{R_d}) \cdot (1 + WSF_{R_d})
\end{equation}
\paragraph{Land footprint}
%Another critical aspect is land use efficiency.
To address land use efficiency, we aim to define its Land-use Usage Effectiveness (LUE). Thus, let us consider that the data center occupies an area $A_d$ and $E^{IT}_d$ is the total IT energy consumption over one year. Thus, we define the LUE as follows:

\begin{equation}
    LUE_d = \frac{A_d}{E^{IT}_d}
\end{equation}
 
Additionally, we define the Capture Loss Factor (CCLF) as the potential loss of carbon capture due to land occupation. Its regional evaluation is denoted with $CCLF_{R_d}$

The on-site land use impact can be quantified as:

\begin{equation}
    E_j \cdot LUE_d \cdot CCLF_{R_d}
\end{equation}

while the off-site land use impact accounting for regional grid dependencies must consider the Energy Land Use Intensity Factor (ELIF), denoted as $ELIF_{R_d}$. Thus we define the off-site land use impact as follows:

\begin{equation}
    E_j \cdot PUE_d \cdot (P_{R_d} \cdot ELIF_{R_d} \cdot CCLF_{R_d})
\end{equation}

\paragraph{E-Waste footprint}
Finally, let us define the E-Waste Intensity (EWI), denoted as $EWI_d$, representing the ratio of annually non-recycled electronic waste to IT energy consumption. The e-waste impact per job can be computed as:

\begin{equation}
    E_j \cdot EWI_d
\end{equation}

{It is worth noting that obtaining accurate data for these metrics can be challenging. Some values, such as PUE or WUE, may be available from company-issued sustainability reports. Others, like WUE, may need to be approximated using indirect estimates or aggregated figures—for example, by combining water withdrawal and energy consumption data. 
Other data, such as land occupation, can be sourced from publicly available datasets, for instance, those provided by websites like datacenters.com \cite{Datacenters_website}.}

\begin{table}[]
    \centering
    \begin{tabular}{c|c|c}
    \hline
        \textbf{Symbol} & \textbf{Description} & \textbf{Unit} \\ \hline
        $R_d$ & Region & -- \\
        $PUE_d$ & Power Usage Effectiveness & --\\ 
        $P_{d}$ & On-site energy consumption& \%\\ 
        $P_{R_d}$ & Off-site energy consumption& \%\\ 
        $CI_{d}$ & On-site Carbon Intensity & $gCO_2/kWh$\\ 
        $CI_{R_d}$ & Off-site Carbon Intensity& $gCO_2/kWh$\\ 
        $WUE_d$ & Water Usage Effectiveness& $l/kWh$\\ 
        $WSF_{R_d}$ & Water Safety Factor& \%\\ 
        $EWIF_{d}$ & Energy Water Intensity Factor & $l/kWh$\\ 
        $LUE_d$&Land Use Effective&$gCO_2/m^2kWh$\\
        $CCLF_{R_d}$ & Carbon Capture Loss Factor & $gCO_2/m^2$\\ 
        $ELIF_{R_d}$ & Energy Land Use Intensity Factor & $m^2/kWh$\\ 
        $EWI_{d}$ & E-Waste Intensity & $g/kWh$\\
        \hline
        
    \end{tabular}
    \caption{Metrics Notation Summary}
    \label{tab:sustainability_profile}
\end{table}

\section{Cloud orchestration}
\label{sec:problem}
In this section, we formally define the problem of job deployment and migration considering sustainability profiles of data centers geographically located in several regions and user sustainability preferences that may derive from the need to adhere to sustainability standards and laws. Data center profiles represent various environmental impact factors and user requirements are prioritized over such factors. 

\begin{figure}
    \centering  
    \includegraphics[width=\linewidth]{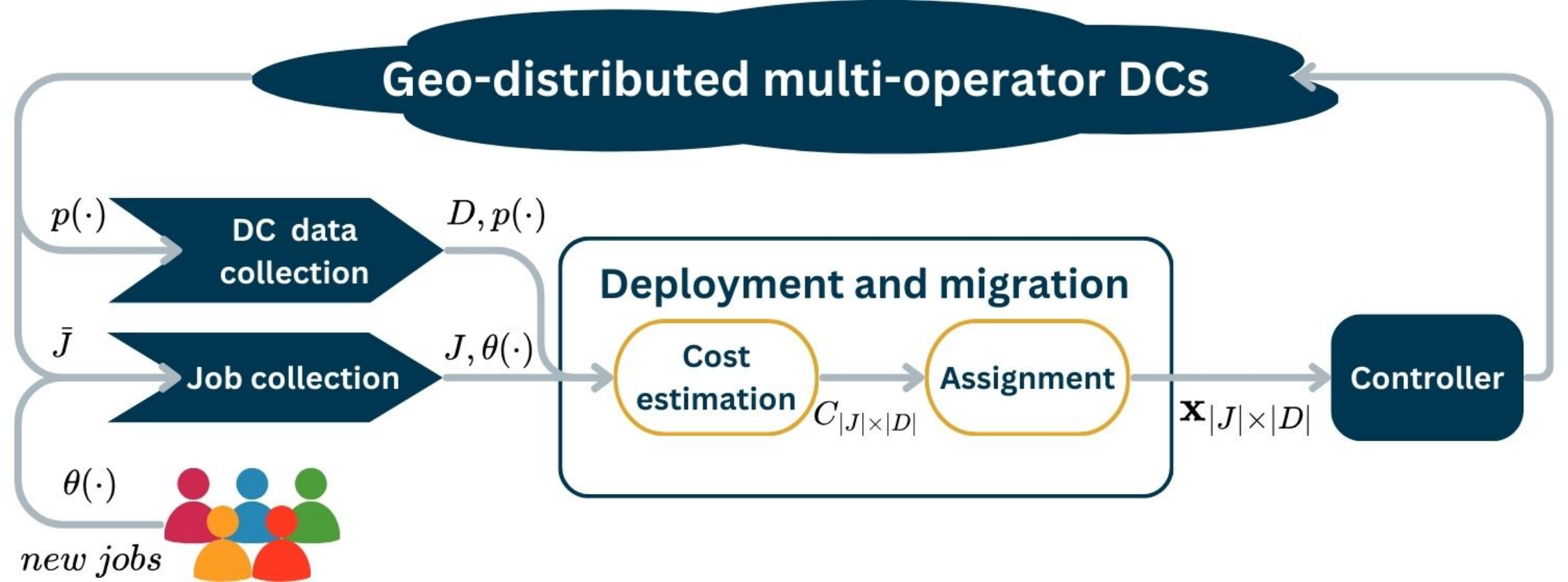}
    \caption{Architecture}
    \label{fig:enter-label}
\end{figure}

\subsection{Deployment and Migration}

\begin{table}[]
    \centering
    \begin{tabular}{c|c}
    \hline
        \textbf{Symbol} & \textbf{Description} \\\hline
        $\mathcal{F}$ & environmental impact factors\\
        $D$ & set of data centers\\
        $J$ & set of job requests\\
        $U$ & set of users\\
        \(u(j)\) & owners of job $j$\\
        $\theta(u)$ & preferences of user $u$\\
        $p(d)$ & sustainability profile of data center $d$\\
        $S^{max}_d$ & {max number of jobs for datacenter $d$}\\
        $d_{prev}(j)$ & data center in which job $j$ is deployed\\
        $\mathds{1}(j)$ & indicator if job $j$ is already deployed\\
        $x_{j,d}$ & assignment decision variable\\\hline
    \end{tabular}
    \caption{Problem  Notation}
    \label{tab:notation}
\end{table}
Let \(\mathcal{F}  = \{1, ..., F\}\)
be the set of environmental impact factors, $J$ the set of job requests, $U$ the set of users, and $D$ the set of Data Centers. Moreover, let us consider the following constant functions: \(u : J \to U\), which is the function providing the owners of each job; \(\theta : U \to \mathbb{R}^F\), which is the function that provides the scores to prioritize factors of interest to meet the user requirements; \(p: D \to \mathbb{R}^F \), which is the function that provides the data centers environmental profiles; \(d_{prev}: \bar{J} \to D\) where \(\bar{J} \subseteq J\) is the subset of already deployed jobs and the function \(d_{prev}\) provides the data center on which each job is deployed; and \(\mathds{1}: J \to \{0,1\} \) returns 1 if its argument belongs to \(\bar{J}\), 0 otherwise.

Finally, let us define a binary decision variable for each \(j \in J\) and \(d \in D\) which indicates whether the  job $j$ is deployed on the data center $d$. Namely, \(x_{j,d} = 1\) if the job $j$ is deployed on the data center $d$ and \(x_{j,d} = 0\) otherwise.
Now, let us define the cost of the decision \(x_{j,d}\) as follow: 
\(C(j,d) = p(d)^{T}\theta(u(j)) \).

Finally, we can define the optimization problem:
\begin{align}
\min \quad & \sum_{j\in J} \sum_{d \in D} C(j,d) \cdot x_{j,d}\\
\text{s.t.} \quad & \sum_{d \in D} x_{j,d} 
    = 1& \forall j \in J \label{eq:assignment}\\
    &\sum_{j\in J\ s.t.\ d_{prev}(j) \neq d}x_{j,d} \le S^{max}_{d}& \forall d\in D\label{eq:capacity}\\
    &\frac{\mathds{1}(j)\cdot C(j,d)}{C(j,d_{prev}(j))} x_{j,d}\le(1-\alpha) & \forall d\in D -\{p(j)\}, j\in J\label{eq:migration_trigger}\\
    &x_{j,d}  \in \{0,1\}& \forall d\in D, j\in J
\end{align}

The objective function minimizes the total environmental impact by considering user-specific factor prioritization and data center sustainability profiles. 
The constraint of Equation \ref{eq:assignment} ensures that each job is deployed in exactly one data center.  
Equation \ref{eq:capacity} limits the number of jobs that can be assigned to each data center for capacity reasons, {where $S^{max}_{d}$ is the maximum allowed number of jobs.}
Finally, equation \ref{eq:migration_trigger} introduces a migration trigger based on an improvement threshold \( \alpha \), ensuring that a job is migrated only if its new deployment location significantly reduces environmental impact, e.g. $\alpha = 0.1$ means that performing the migration brings a 10\% improvement.  It is easy to prove that the generalized assignment problem is reducible to our problem, which is therefore NP-hard.

The optimization is meant to occur periodically{, i.e. every $\Delta_t$ seconds,} updating the sustainability profiles and the job status and queue{, as detailed in Algorithm \ref{alg:iterative_scheduling}.}
To ensure the feasibility of the problem, we assume that the maximum load of jobs in the system is always smaller than the total capacity.

{

\begin{algorithm}
\caption{Iterative MIP-Based Scheduling Procedure}
\begin{algorithmic}[1]
\State \textbf{Input: $D$ data centers} 
\For{$t = 0, \Delta_t, 2\Delta_t, ... H\Delta_t$}
    \State Update $\bar{J}$ with running jobs at time $t$
    \State Get sustainability profiles $p_d$ for each $d \in D$
    \State Update $J$ with new jobs
    \State Instantiate MIP model for time step $t$
    \State Solve the model to obtain solution $x_t$
    \State Perform scheduling using solution $x_t$
\EndFor
\State \textbf{Output:} Final schedule over horizon $H$
\end{algorithmic}
\end{algorithm}
\label{alg:iterative_scheduling}
}

\section{Experiments}
{To evaluate our approach, we conduct two simple case studies.}
We leverage real data aggregating several sources, e.g. the energy source mix and the regional and the data center-specific environmental parameters. However, some data has been randomly generated, e.g. the users' requests.
In this section, we first present our experimental settings, then we present the results of three experiments. The first experiment aims to compare our approach with single-factor optimization baselines. The second experiment is meant to highlight the potential benefit of enabling migration. The last experiment is a discussion/consideration on existing empirical models to estimate WUE values and the reported data we use throughout our experiments.

\label{sec:experiments}
\subsection{Experimental Setting}

Our experimental setup aims to evaluate our approach considering a real-world geo-distributed cloud environment. We consider multiple data centers from different cloud providers.

\paragraph{Data Sources} 
We selected 24 data centers across different cloud providers: 13 by Meta, 7 by Google, 4 by Azure, and 5 by AWS.  {In the first scenario we considered only Meta data centers. The second scenario considers data centers from the could providers Meta, Google, and Azure.}
For each data center, we exploited the data provided in section \ref{sec:profiles} to compute the sustainability profiles of data centers.
Due to the lack of homogeneous and fine-grained data, estimations are necessary to derive comparable metrics from insights into the sustainability of each data center.
For Meta data centers we estimated the IT power consumption by applying the global annual PUE to the annual data center-specific facility power consumption reported by the company.
The annual IT power consumption for Google and Azure data centers is estimated using the reported data center-specific PUE {and considering that a hyperscale data center is estimated to consume at least 100 MW \cite{iea2024aiboom}, which yields a consumption of at least 876.000 MWh.
Estimating the power consumption of data centers is a complex task. Global and regional estimates are based on utilization estimates, hardware specifications, and marked analysis to quantify the amount of hardware shipped to a certain region \cite{iea2024energyai,eu2024cloud,lbnl2024datacenters}. Thus, it is difficult to come up with an estimate with a finer granularity. }
Waste management data is unavailable to compare data centers, so we excluded the related factor from the case study.
To the best of our knowledge, also information about data centers' on-site power grid was unavailable. Thus, our case study assumes that all the power is provided by the regional grid.

To gather regional data we relied on:
\begin{itemize}
    \item Electricity Maps API \cite{ElectricityMaps_API} to retrieve electricity grid data which allowed us to get the grid's power source mix.
    \item Our World In Data to retrieve intensity factors per power source. Namely, the Carbon Intensity \cite{OurWorldInData_Safest_Energy_Sources} for parameters $CI_{R_d}$ and the Energy Land Use Intensity Factor \cite{OurWorldInData_Land_Use_Energy_Sources} for parameters $ELIF_{R_d}$.
    Also the parameters $WSF_{R_d}$ are based on data from Our World in Data \cite{OurWorldInData_Freshwater} to represent freshwater withdrawals as a share of internal resources.
    \item The National Renewable Energy Laboratory (NREL) report \cite{NREL_50900} to assess water demand of a certain energy mix and assign values to the parameter $EWIF_{R_d}$.
    \item {Estimates for $CCLF_{R_d}$ at state or regional level are taken from various sources (Ireland \cite{cclfIR}, Denmark \cite{totcarbremDK,forestryDK}, Sweden \cite{cclfSW}, Australia \cite{cclfAU}, US \cite{cclfUSr1,cclfUSr2,cclfUSr3,cclfUSr4,cclfUSr5,cclfUSr6,cclfUSr8,cclfUSr9,cclfUSr10}). }
\end{itemize}

\paragraph{Time Horizon}  
The experiments cover a {72}-hour period from {\texttt{2025-05-12 00:00 UTC} to \texttt{2025-05-14 00:00 UTC}}. Every hour the model is instantiated with updated parameters and solved.

\paragraph{Job Request Generation}  
Job requests were randomly generated we manually defined three users with different sustainability needs. 
Each user has a random amount of job requests, which were randomly generated. For each job the expected hourly power consumption is drawn from a uniform distribution between 0.5 kW and 10 kW; the expected lifetime is drawn from a uniform distribution between 1 and 5 hours. The job arrival process is modeled as a Poisson process with \(\lambda = 10\) jobs per hour. {Note that in the first scenario, users can be interpreted as the operator having different preferences on different jobs.} 

To ensure MIP feasibility, each data center was limited to a maximum of 5 concurrent jobs. {This limit is hand-picked for the sake of demonstrating the potential of our framework and does not reflect real large-scale data center constraints.}

{Each simulation is repeated for 10  different randomly generated sets of job requests.}

\paragraph{Baseline Comparisons}  
To assess the effectiveness of our preference-based optimization approach, we compare it against three baseline models, each optimizing for only one sustainability factor. Namely, \textbf{\carbonopt}, \textbf{\wateropt}, and \textbf{\landuseopt}.

\paragraph{Implementation Details}  
The experiments have been implemented in Python and the code is publicly available \cite{github_repo}.  
The optimization model is implemented using the PuLP \cite{pulp} library and Gurobi \cite{gurobi} is used as solver.

\subsection{Results}
\paragraph{Baseline Comparison}

\begin{figure*}
\centering
    \begin{subfigure}[b]{0.45\textwidth}
    \centering
    \includegraphics[width=\linewidth]{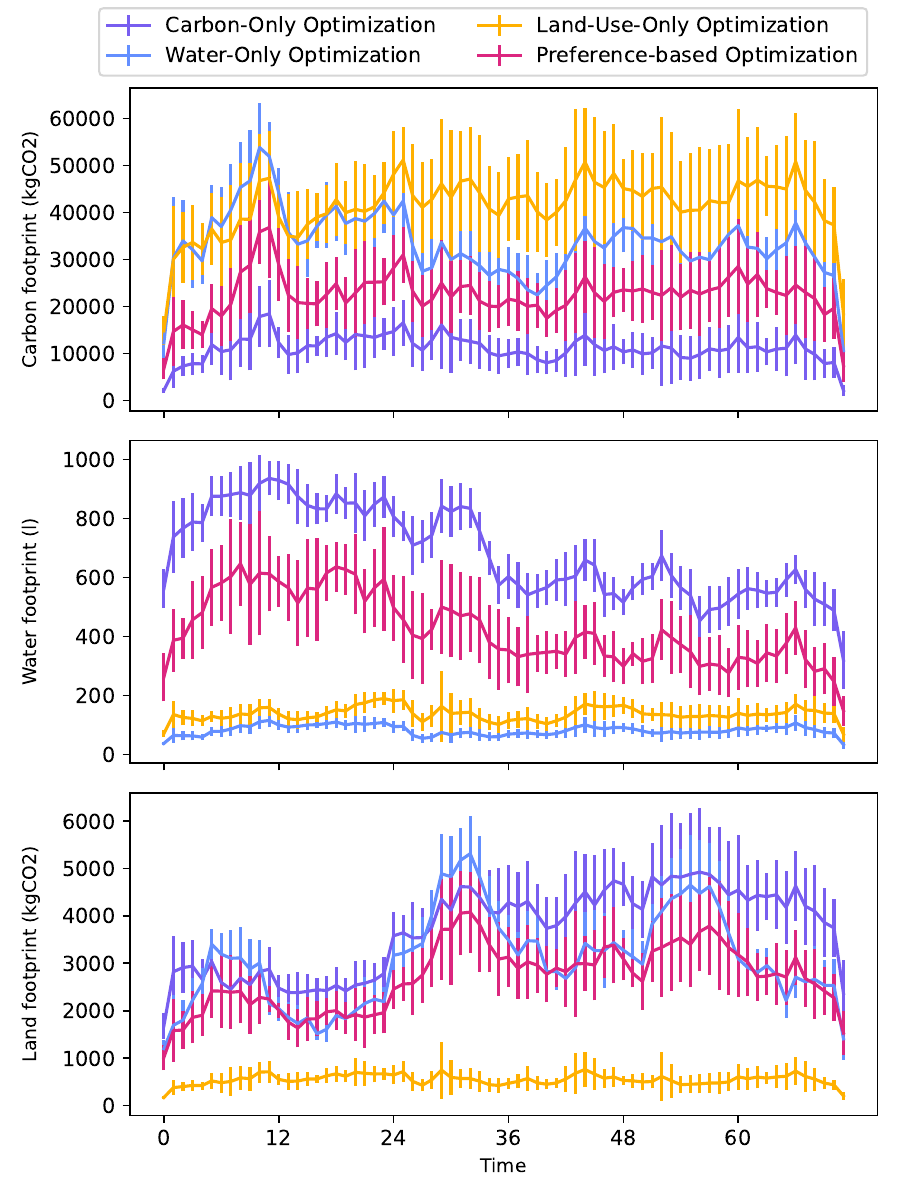}
    \caption{Meta scenario.}
    \label{fig:meta_comparison}
    \end{subfigure}
    \begin{subfigure}[b]{0.45\textwidth}
    \centering
    \includegraphics[width=\linewidth]{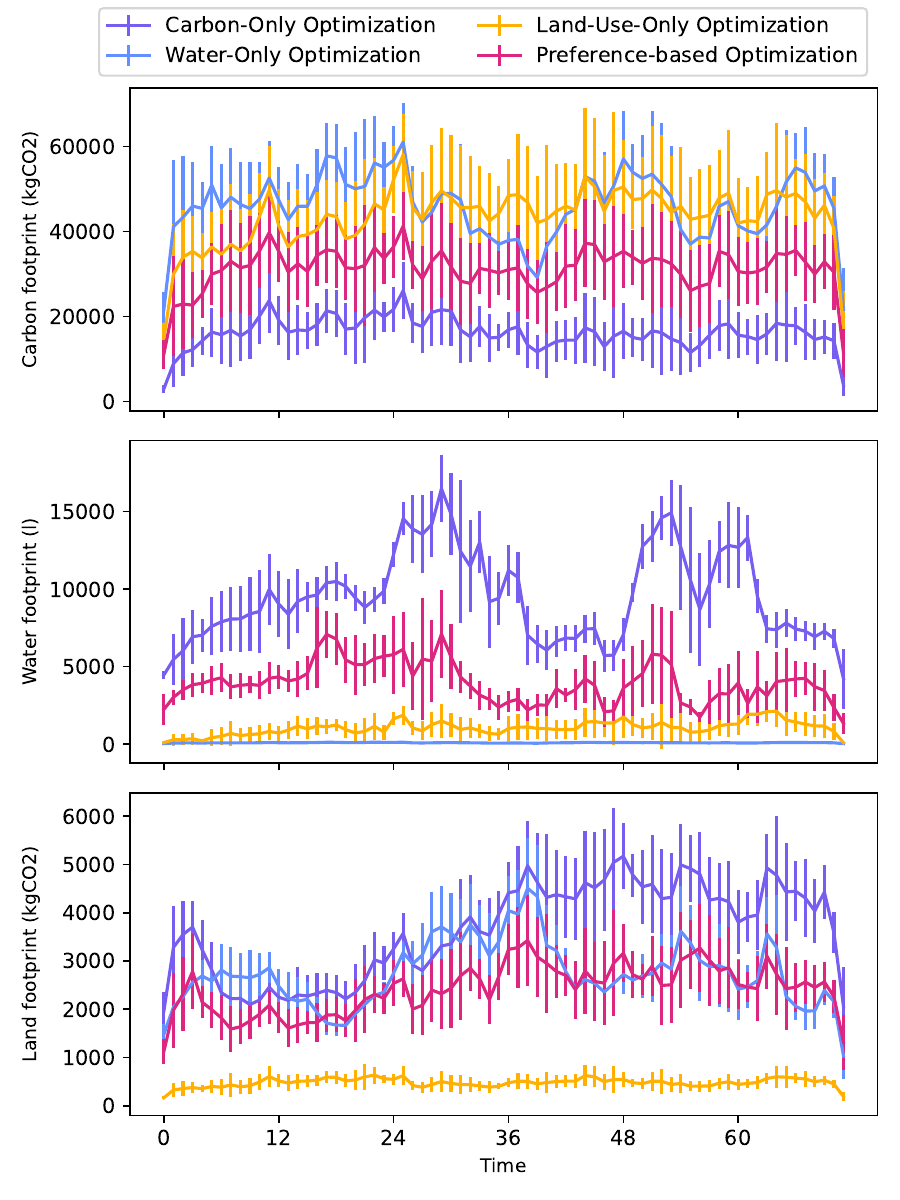}
    \caption{Cloud platform scenario.}
    \label{fig:cp_comparison}   
    \end{subfigure}
    \caption{Baseline Comparison}
\end{figure*}

\begin{figure}
    \centering
    \includegraphics[width=\linewidth]{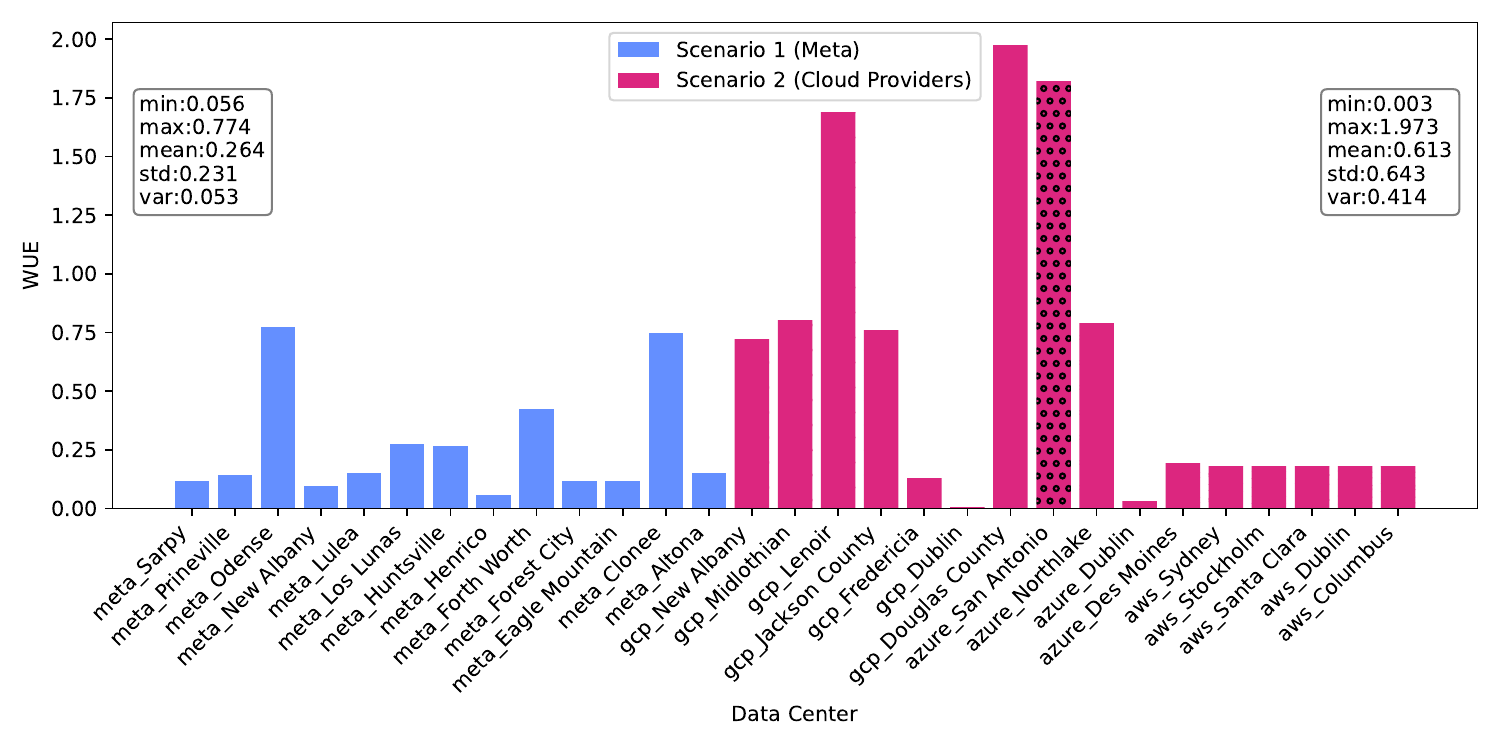}
    \caption{WUE values of Data Centers included in the experiments. Naming rule: \texttt{provider\_location}.}
    \label{fig:dcs_wue}
\end{figure}

Figures \ref{fig:meta_comparison} and \ref{fig:cp_comparison} present the comparative analysis of different optimization approaches in terms of carbon footprint, water footprint, and land use over a {72}-hour period. The impact on sustainability metrics varies significantly depending on the optimization strategy.
The plot on the top of the figures illustrates carbon emissions, where the \carbonopt approach (green) achieves the lowest emissions, as expected. However, this comes at the cost of an increased water footprint, as shown in the plot in the middle, where \carbonopt exhibits the highest water footprint due to the selection of water-intensive data centers. Similarly, the \wateropt strategy (blue) minimizes water footprint but leads to higher carbon emissions (middle graph), demonstrating the necessity to trade-off between carbon and water footprint.
The plot on the bottom depicts land use, where the \landuseopt approach (yellow) minimizes land usage but results in elevated carbon emissions. In contrast, in this scenario in which each user has a different preference, our \preferencebased (red) achieves a balanced trade-off across all sustainability factors, reducing extreme variations and ensuring a more holistic optimization. \textit{This highlights the advantage of an approach that incorporates user-defined trade-off preferences rather than focusing on a single metric}.
{The two scenarios present similar results, except the fact that when we consider cloud providers there is more variability in water footprint, due to the fact that there is more variance in the WUE values, as shown in Figure \ref{fig:dcs_wue}.}

%---------------

\paragraph{Migration}

\begin{figure*}
\centering
    \begin{subfigure}[b]{0.45\textwidth}
    \centering
    \includegraphics[width=\textwidth]{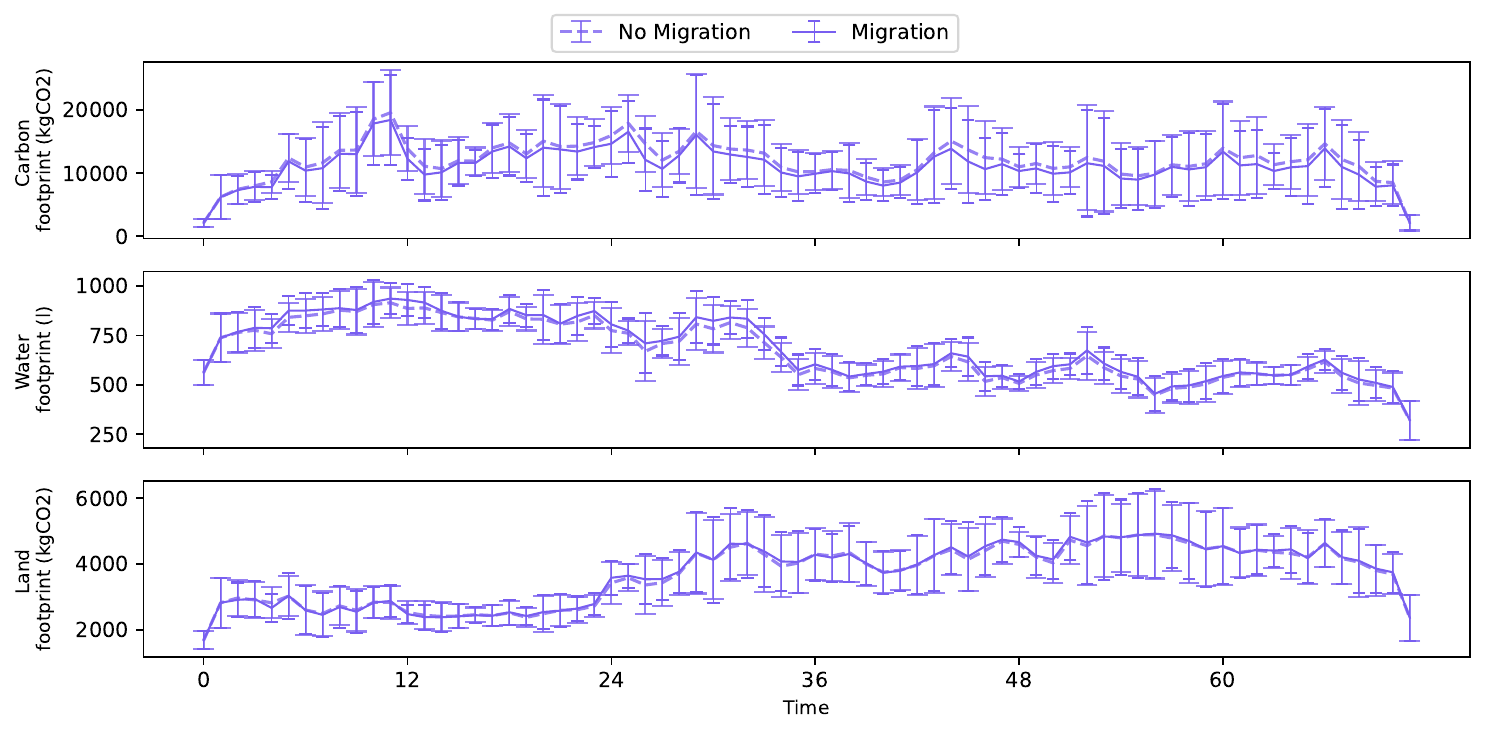}
    \caption{\label{fig:meta_carbon_opt}\carbonopt}
    \end{subfigure}
\quad
    \begin{subfigure}[b]{0.45\textwidth}
    \centering
    \includegraphics[width=\textwidth]{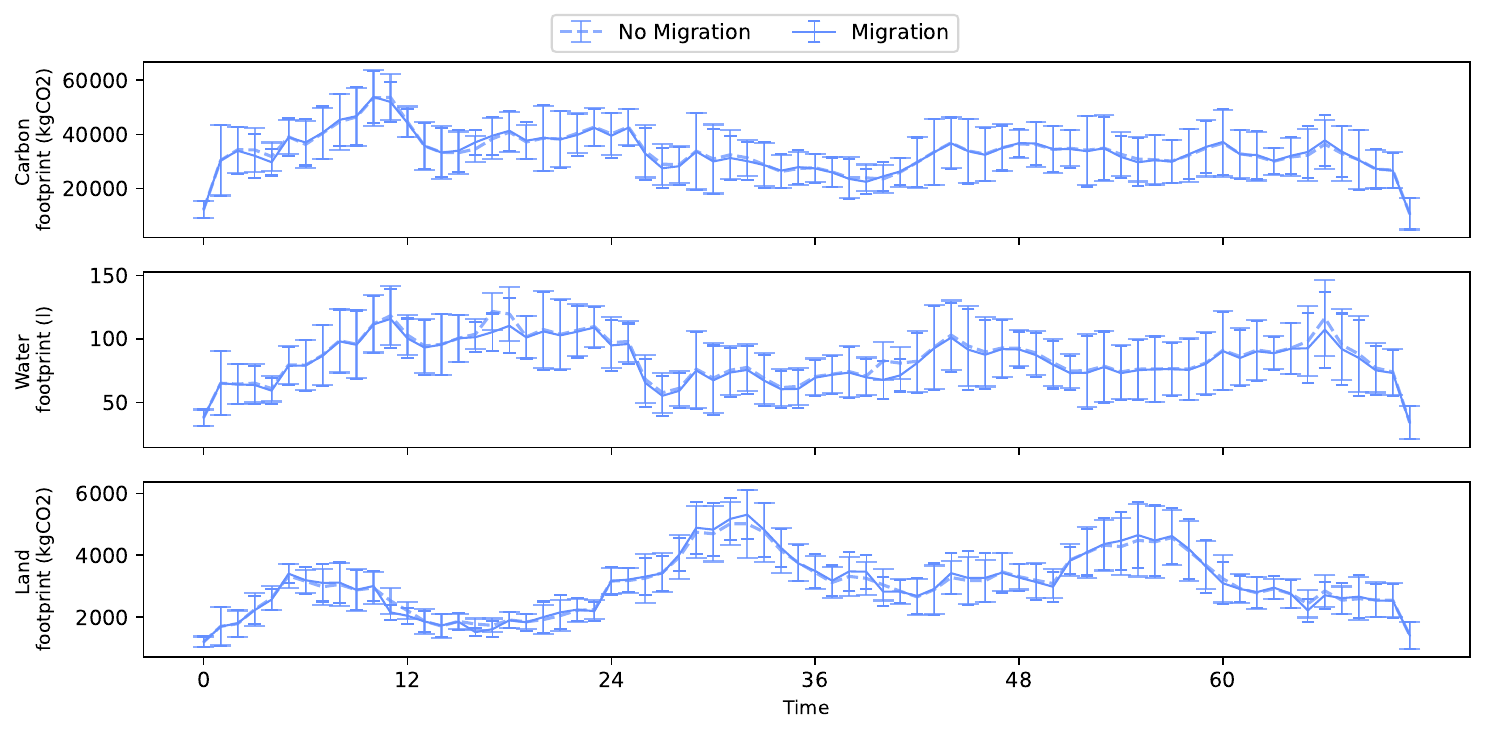}
    \caption{\label{fig:meta_water_opt}\wateropt}
    \end{subfigure}
\\
    \begin{subfigure}[b]{0.45\textwidth}
    \centering
    \includegraphics[width=\textwidth]{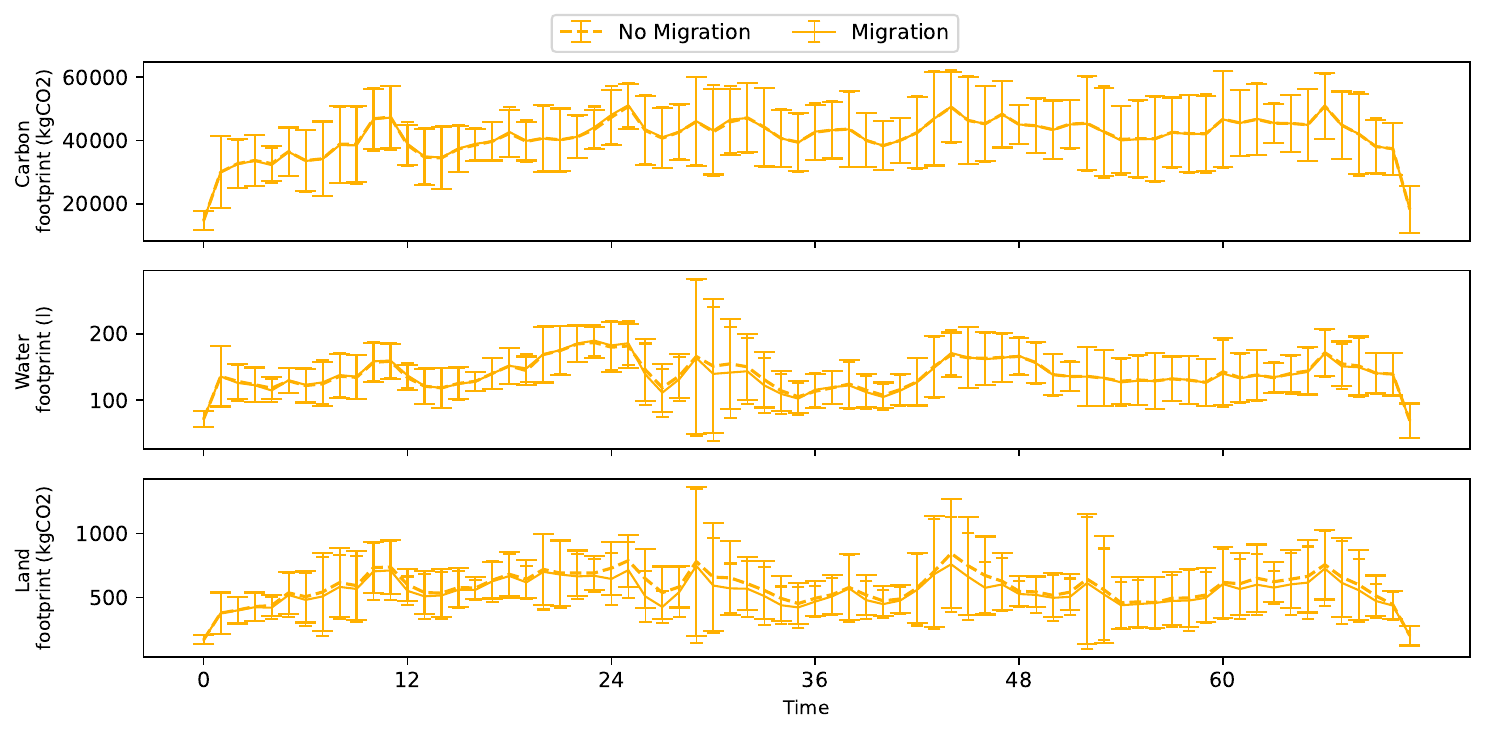}
    \caption{\label{fig:meta_land_use_opt} \landuseopt}
    \end{subfigure}
\quad
\begin{subfigure}[b]{0.45\textwidth}
    \centering
    \includegraphics[width=\textwidth]{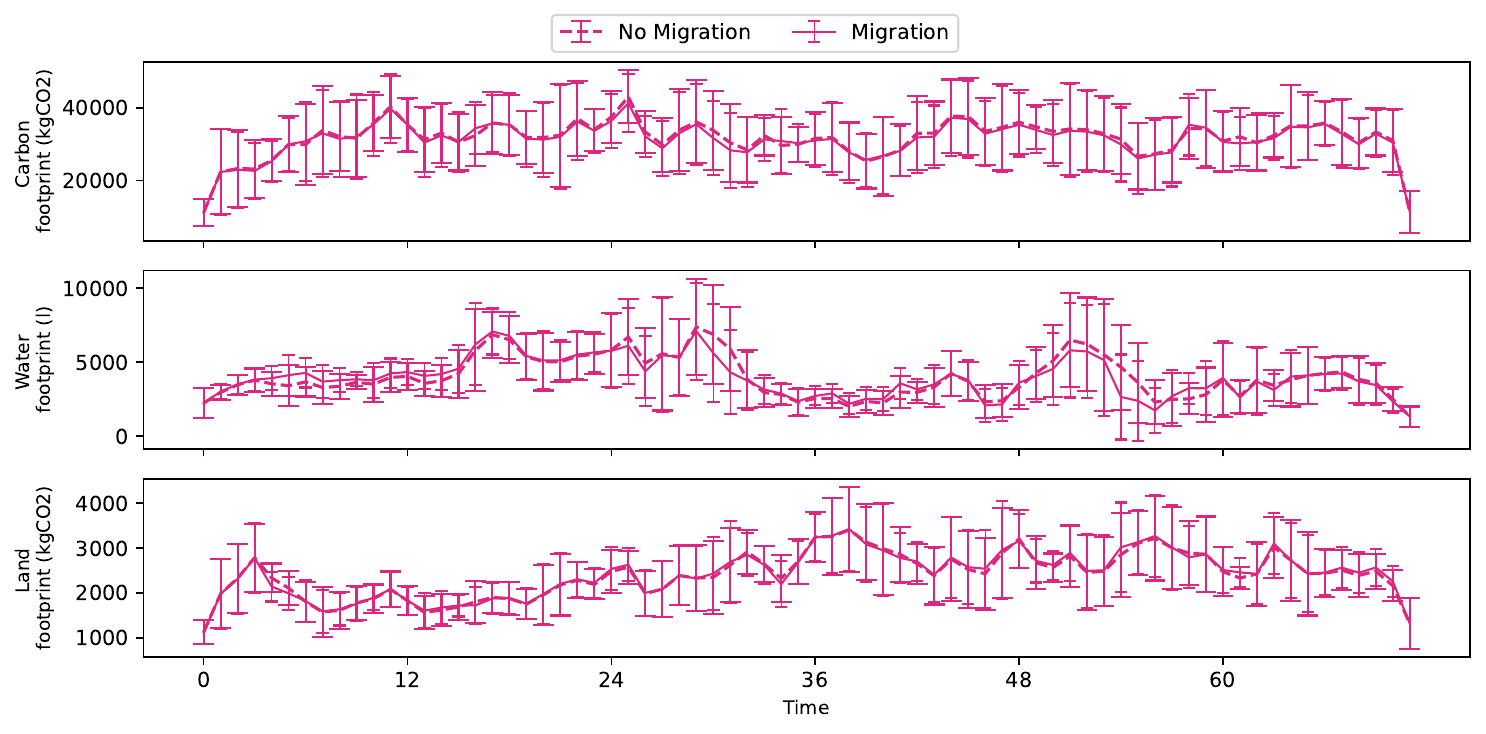}
    \caption{\label{fig:meta_preference_based_opt}\preferencebased}
    \end{subfigure}

\caption{Meta experiment.
}
\label{fig:meta_migr}
\end{figure*}

% ---------

\begin{figure*}
\centering
    \begin{subfigure}[b]{0.45\textwidth}
    \centering
    \includegraphics[width=\textwidth]{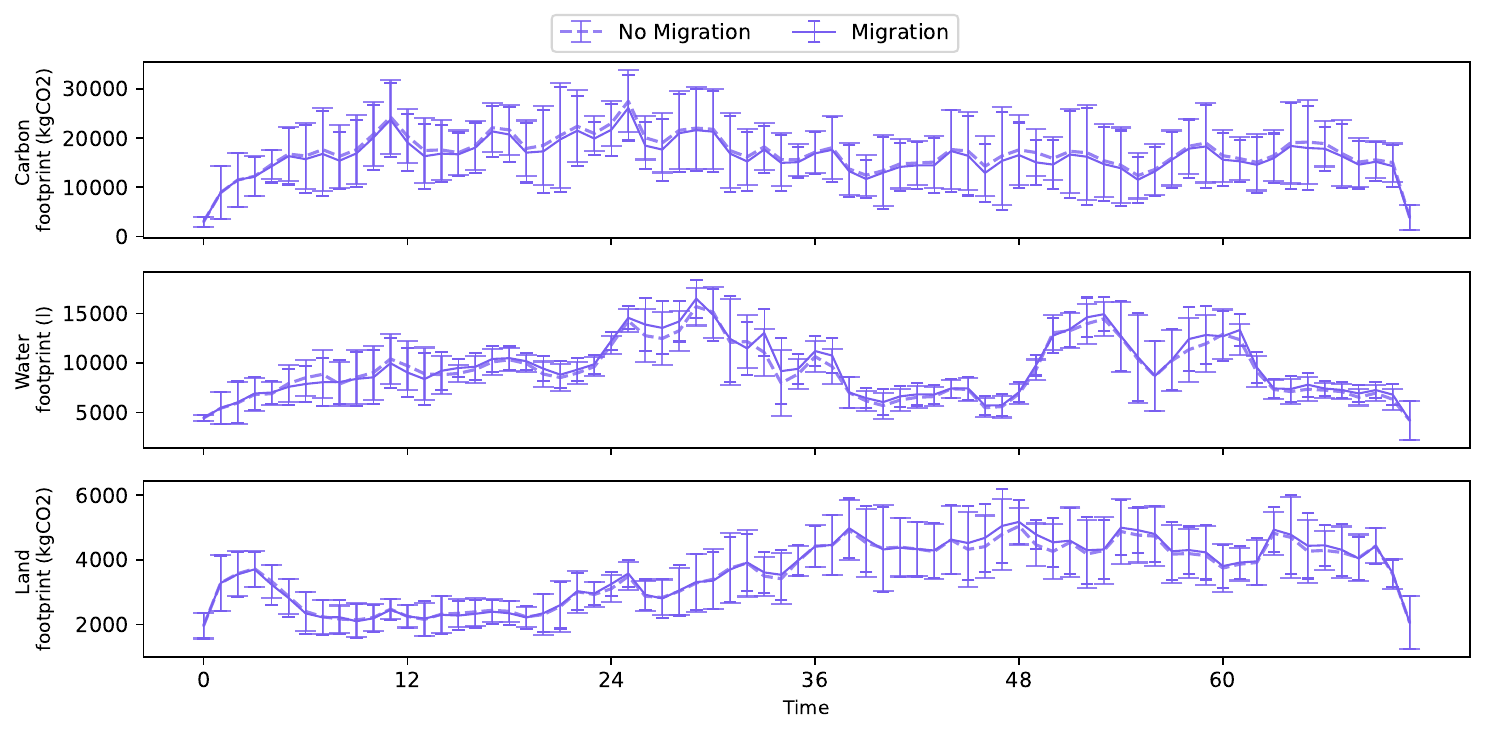}
    \caption{\label{fig:cp_carbon_opt} \textbf{\carbonopt}}
    \end{subfigure}
\quad
    \begin{subfigure}[b]{0.45\textwidth}
    \centering
    \includegraphics[width=\textwidth]{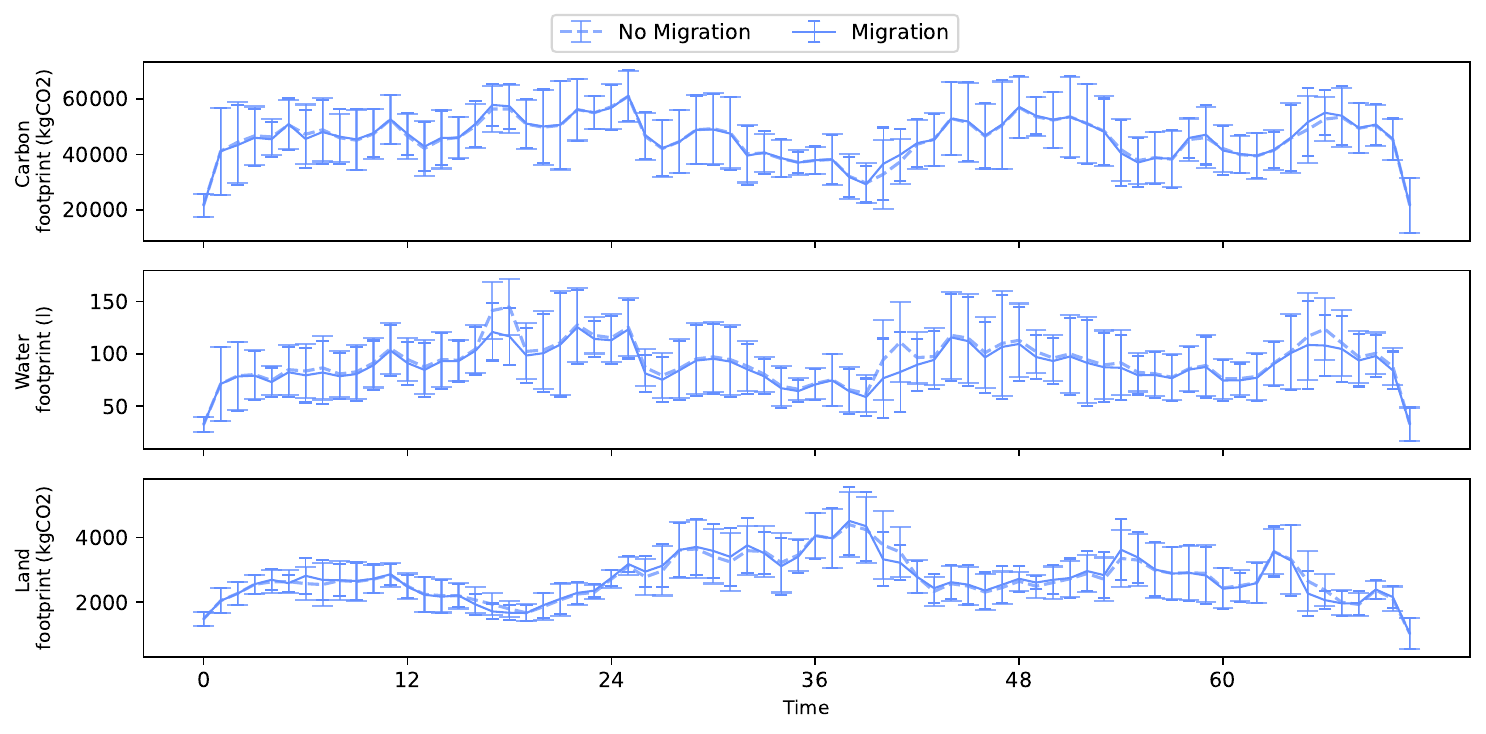}
    \caption{\label{fig:cp_water_opt} \textbf{\wateropt}}
    \end{subfigure}
\\
    \begin{subfigure}[b]{0.45\textwidth}
    \centering
    \includegraphics[width=\textwidth]{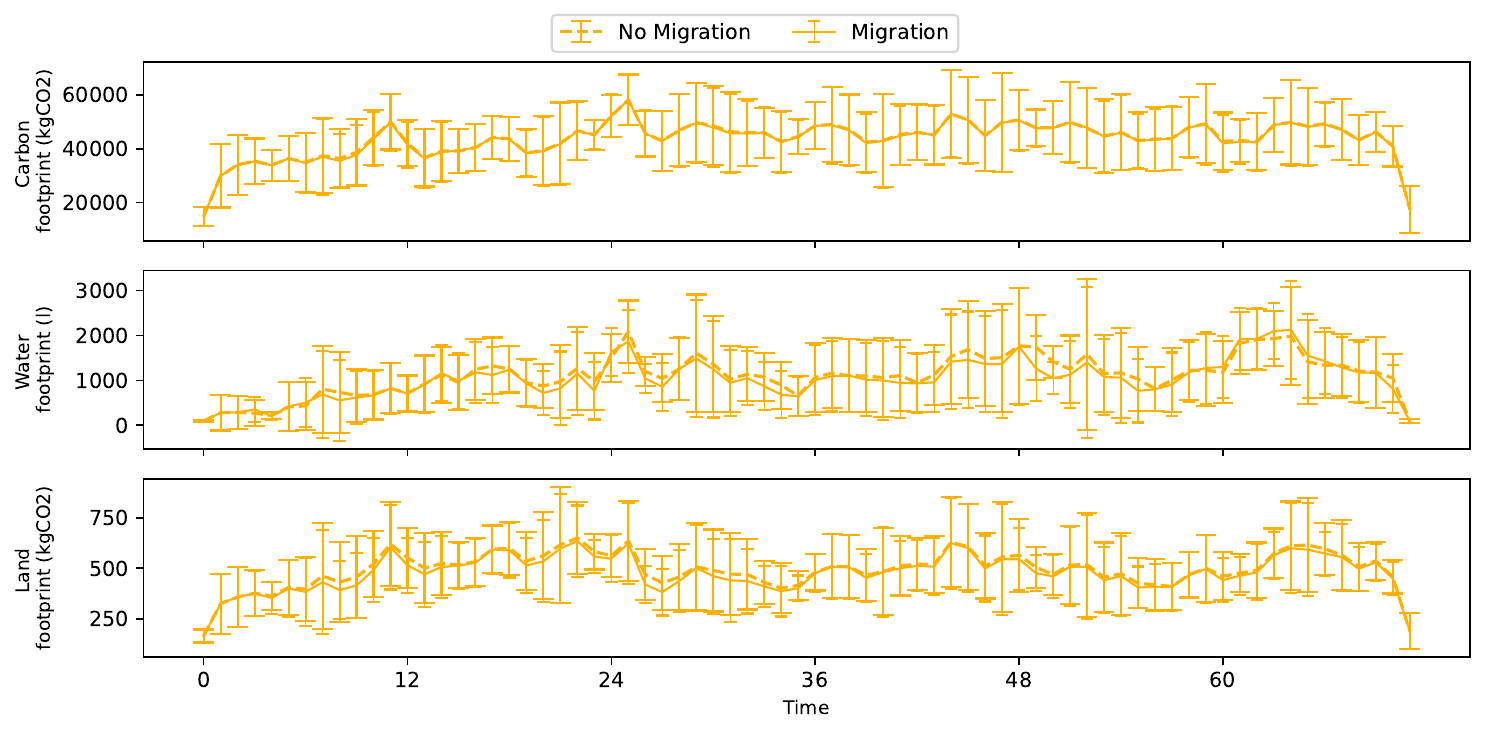}
    \caption{\label{fig:cp_land_use_opt}\textbf{\landuseopt}}
    \end{subfigure}
\quad
\begin{subfigure}[b]{0.45\textwidth}
    \centering
    \includegraphics[width=\textwidth]{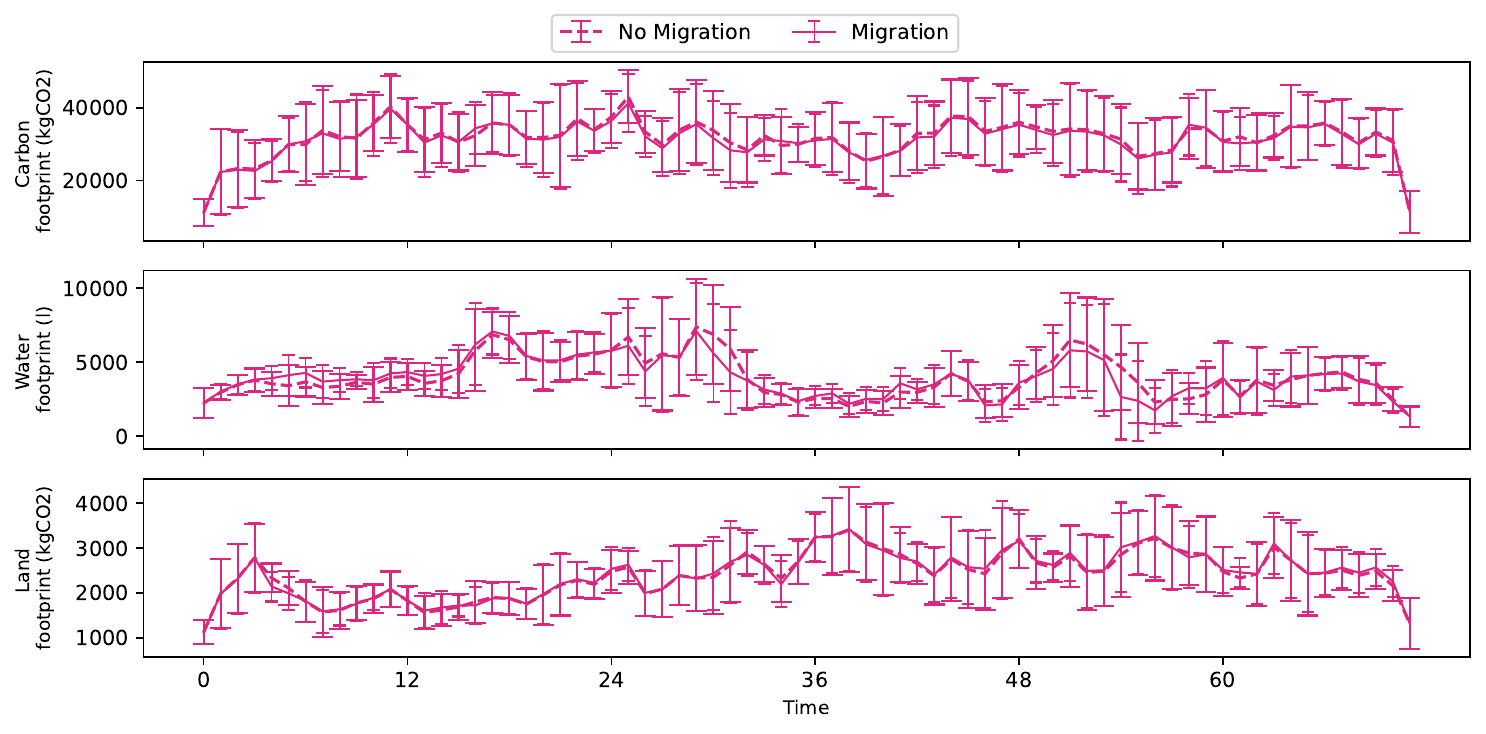}
    \caption{\label{fig:cp_preference_based_opt} \textbf{\preferencebased}}
    \end{subfigure}

\caption{Could platforms experiment}
\label{fig:cp_migr}
\end{figure*}

{Figures \ref{fig:meta_migr} and \ref{fig:cp_migr} show for each approach the performance comparison of enabling migration or not, respectively for Meta and cloud providers. }

Considering the single factor optimization it is clear that migration results in a reduction of the impact of such factor both for Meta (Figures \ref{fig:meta_carbon_opt} top, \ref{fig:meta_water_opt} middle, \ref{fig:meta_land_use_opt} bottom) and for the cloud providers (Figures \ref{fig:cp_carbon_opt} top, \ref{fig:cp_water_opt} middle, \ref{fig:cp_land_use_opt} bottom). 
The \carbonopt baseline results in a slight improvement in carbon emission. However, those migrations also slightly increase the water footprint (top, middle graphs in Figures \ref{fig:meta_carbon_opt} and \ref{fig:cp_carbon_opt}).  

On the other hand, with \wateropt it is not true that migrations reducing water footprint always results in a rise of carbon emissions, {for example in Meta's scenario we can observe peak reductions from hour 17 to hour 19 which result in a slight increase in carbon footprint and a peak from hour 40 to hour 42 which does not result in any carbon footprint increase Figure \ref{fig:meta_water_opt}.
For what concerns Meta's scenario, from Figure \ref{fig:meta_land_use_opt} it is possible to observe that the improvement obtained in land use with \landuseopt are negligible as the side effects on the other metrics. On the other hand, for the other scenario, the effects on carbon emissions of these improvements are negligible, but it seems to result in a reduction of water footprint too, as shown in Figure \ref{fig:cp_preference_based_opt}.
}

\paragraph{Declared WUE vs. Estimated WUE}
\begin{figure*}
\centering
    \begin{subfigure}{0.45\textwidth}
    \centering
    \includegraphics[width=\textwidth]{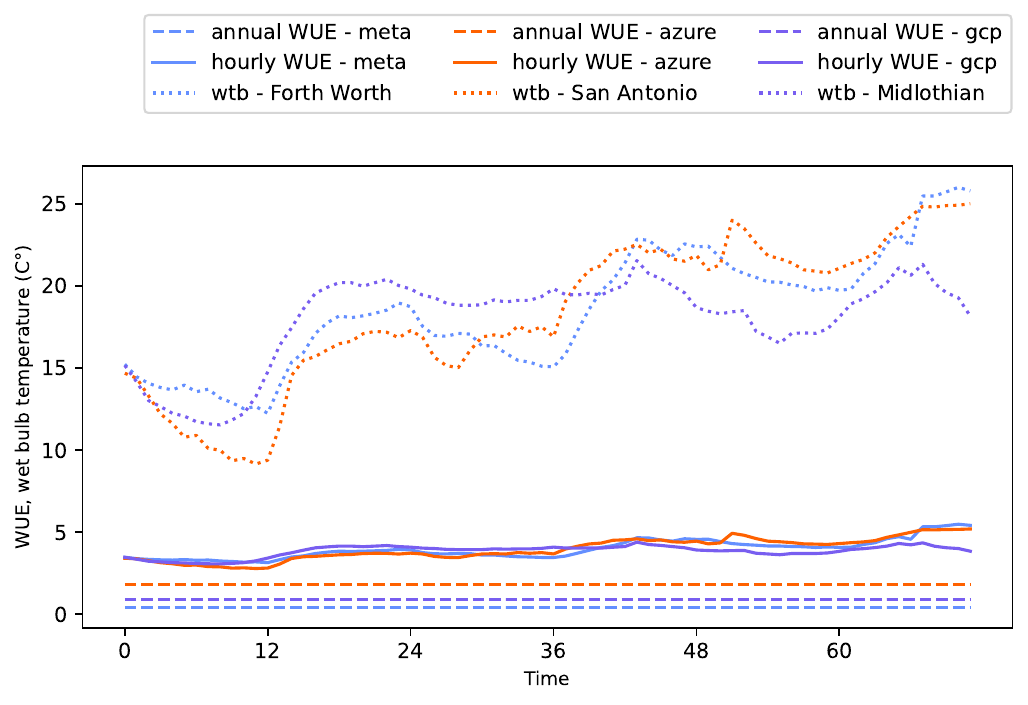}
    \caption{\label{fig:wue_texas} Texas}
    \end{subfigure}
\quad
    \begin{subfigure}{0.45\textwidth}
    \centering
    \includegraphics[width=\textwidth]{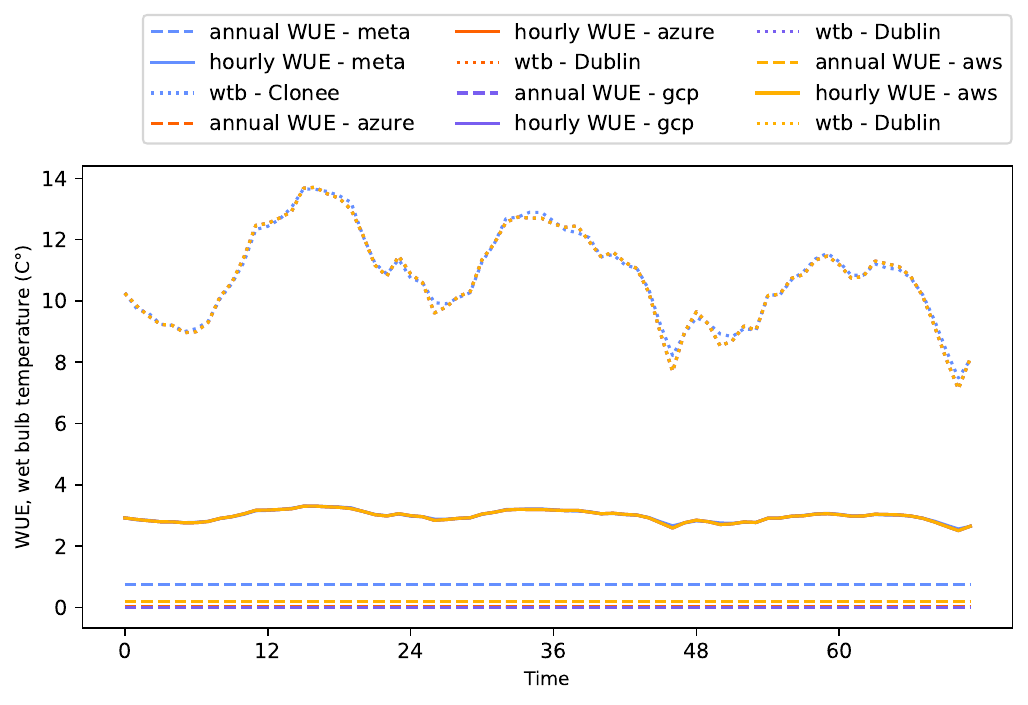}
    \caption{\label{fig:wue_ireland} Ireland}
    \end{subfigure}
\caption{Reported and estimated WUE comparison}
\end{figure*}
In our experiments, we used the annual WUE value reported by companies or approximated from other reported values. However, a formula allowing for a time-varying estimation of WUE values has been proposed by Islam et al. \cite{7152842}. The authors used data fitting to obtain the following empirical model for estimating WUE:
\[
\frac{s}{s-1} {6\times 10^5 T^3_w - 0.01 \times 10^2 T^2_w +0.61 T_w - 10.40}
\]
This model depends on the cycles of concentrations \cite{cycleofconcentrations} $s$ and the outside wet bulb temperature (in Fahrenheit) $T_w$. 
This formulation has already been used in other works \cite{li2023making, jiang2025waterwise}.

We implemented such a model by estimating the wet bulb temperature from the temperature and the dew point temperature from meteorological data \cite{VisualCrossing_API}. We set $s$ to 10, which is the parameter that resulted in smaller WUE values in the experiments conducted by \cite{7152842}. Moreover, we consider the WUE to be 0 for the temperatures for which the model returns negative values.

Figures \ref{fig:wue_ireland} and \ref{fig:wue_texas} show WUE values comparing the cloud providers we have considered so far. We selected two states in which all providers have a data center, namely Ireland (Figure \ref{fig:wue_ireland}) and Texas (Figure \ref{fig:wue_texas}).
The annual WUE values, represented with dashed lines, correspond to the values reported by the company or estimated based on available data. The hourly WUE values, shown with solid lines, were computed using the aforementioned empirical model. Additionally, the dotted lines represent the wet-bulb temperatures estimated from the meteorological data of the cities in which the data centers are located. 

Annual WUE reported or estimated values may not fully capture the dynamic of water efficiency of data center. However, it is important to note that the difference between such values and the ones provided by the empirical model data is significant. Thus, further studies should be conducted in this direction.

\section{Related work}
\label{sec:relwork}
Enhancing resource management and operational efficiency, cloud orchestration has fueled a wide body of research \cite{10.1145/3054177}.
Allocating resources, both optimizing energy consumption \cite{10609608, 10169097, 10.1145/2342356.2342398, 7345588} and carbon awareness \cite{10.1145/3634769.3634812, 10.1145/2342356.2342398, 7345588, 10.1145/3604930.3605711, 10305816, piontek2024carbon, 10.1145/3631295.3631396} have received a significant interest. 
Stojkovic et al. \cite{10609608} propose EcoFaaS, an energy management framework, designed for serverless environments, that optimizes overall energy consumption.
Rastegar et al. \cite{10169097} propose an LP relaxation for an energy-aware execution scheduler for serverless service providers, minimization of energy consumption for executing the incoming function chains with specified computational loads and deadlines.
Gao et al. \cite{10.1145/2342356.2342398} provide a scheduler that controls the traffic directed to each data center optimizing a 3way-trade-off between access latency, electricity cost, and carbon footprint.
Zhou et al. \cite{7345588} use
Lyapunov optimization to perform load balancing across geo-distributed data centers, capacity right-sizing (turning off idle servers), and server speed scaling (adjusting the CPU frequency). The objective here is to optimize a 3way-trade-off between electricity cost, SLA (Service-Level Agreement) requirements, and emission reduction budget.
Souza et al. \cite{10.1145/3634769.3634812} propose a provisioner that minimizes both the number of active servers and the associated carbon emission.
Maji et al. \cite{10.1145/3604930.3605711} propose load balancing in VMware’s Avi Global Server Load Balancer, which uses a linear scoring function to select the optimal data center in terms of marginal carbon intensity and the distance between the client and the data center.
Cordingly et al. \cite{10305816} propose a prototype for computing resource aggregation that minimizes the carbon footprint of a serverless application through carbon-aware load distribution.
Piontek et al. \cite{piontek2024carbon} propose a Kubernetes scheduler that shifts non-critical jobs in time so to reduce carbon emissions based on their prediction algorithm.
Chadha et al. \cite{10.1145/3631295.3631396} present GreenCourier, a Kubernetes scheduler designed to reduce carbon emissions associated with serverless functions scheduled across geographically distributed regions.
Recent research has started to address the environmental impact of cloud and AI workloads beyond just carbon emissions. Li et al. \cite{li2023making} highlight the hidden water footprint of AI models, uncovering the significant amounts of freshwater consumed for both cooling and electricity generation. Their study estimates that training large AI models results in hundreds of thousands of liters of direct water consumption.
Jiang et al. \cite{jiang2025waterwise} introduce WaterWise, a job scheduling framework that co-optimizes carbon and water sustainability for geographically distributed data centers by employing a Mixed Integer Linear Programming (MILP)-based scheduler, which dynamically shifts workloads to regions with optimal carbon and water conditions while considering delay tolerance constraints.
This work aligns with ours, which also emphasizes the need for multi-dimensional sustainability metrics when optimizing cloud workload allocation, highlighting the need for comprehensive sustainability-aware orchestration strategies.

Differently from previous works, we aim to consider sustainability as a whole, accounting for more than energy consumption and the related GHG emissions. We aim to propose a holistic approach that assesses and optimizes the overall impact.

\section{Conclusions and future works}
\label{sec:conclusions}
\paragraph{{Summary of contributions}}
In this work, we addressed the environmental impact of cloud computing by proposing an environmentally conscious cloud orchestrator. Our approach is grounded in a thorough analysis of sustainability reports, from which we extracted meaningful environmental indicators to inform decision-making. Based on these insights, we defined sustainability profiles and formalized metrics to assess the environmental footprint of data centers. Leveraging these metrics, we introduced an optimization-based orchestrator that assigns and migrates jobs while minimizing their environmental impact, all while considering users' sustainability preferences. To evaluate our approach, we conducted {two simulative case studies, demonstrating its potentialities} in balancing the environmental footprint compared to baseline strategies that optimize only a single factor.

\paragraph{{Threats to validity}}
{
Several factors may affect the validity and generalizability of our results.
First, the model does not incorporate uncertainties in grid conditions, such as fluctuations in renewable energy availability, electricity prices, or grid congestion, which can significantly influence energy-aware workload placement.
Second, the study relies on synthetic workloads rather than real-world traces, which may not capture the full variability and complexity of production environments.
Third, we lack access to proprietary data from actual data center operators. This limits the realism of certain assumptions and may affect the applicability of the results.
Finally, we assume that job migration is always technically and legally feasible. In practice, this may not hold due to infrastructure-level heterogeneity or legal restrictions, such as data residency and compliance requirements.
}

\paragraph{{Limitations and future works}} 
Despite the promising results, we recognize several areas for improvement. An extensive experimental evaluation is necessary to ensure statistical confidence in the findings. {Such an evaluation should also include more realistic assumptions and settings.}
Furthermore, incorporating life cycle assessment (LCA) measurements into sustainability profiles could provide a more comprehensive evaluation of the environmental impact, including the construction and dismantling phases. 
Addressing these limitations presents exciting opportunities for future research aimed at further improving sustainability-aware cloud orchestration.

\bibliographystyle{ACM-Reference-Format}
\bibliography{references}

\end{document}